\documentstyle[12pt,a4,epsf]{article}
\newcommand{\be}{\begin{equation}\label}
\newcommand{\ee}{\end{equation}}
\newcommand{\bea}{\begin{eqnarray}\label}
\newcommand{\eea}{\end{eqnarray}}

\newcommand{\ul}{\underline}
\newcommand{\sgn}{\mbox{sign}}
\begin{document}
\title{Analytical Approach for Piecewise Linear\\
       Coupled Map Lattices}
\author{Wolfram Just\thanks{e--mail: wolfram@mpipks-dresden.mpg.de}\\
        Max--Planck--Institute for Physics of Complex Systems\\
        N\"othnitzer Stra\ss e 38\\
        D--01187 Dresden\\
        Germany}
\date{October 30, 1997}
\maketitle
\begin{abstract}
A simple construction is presented, which generalises piecewise linear 
one--dimensional Markov maps to an arbitrary number of dimensions. The
corresponding coupled map lattice, known as a simplicial mapping in the
mathematical literature, allows for an analytical investigation. Especially
the spin Hamiltonian, which is generated by the symbolic dynamics 
is accessible. {  As an example a formal relation between a globally coupled 
system and an Ising mean field model is established. The
phase transition in the limit of infinite system size is
analysed and analytical} results are compared with numerical simulations.
\end{abstract}
\hspace*{7mm} 
\parbox{135mm}{
\begin{tabular}{ll}
PACS No.: & 05.45.+b\\
Keywords: &\parbox[t]{95mm}{Phase transition, Simplicial mapping, 
Mean field Ising model, Symbolic dynamics}\\
Running title: & Analytical Approach for Coupled Map Lattices \\
Submitted to: & J.~Stat.~Phys.
\end{tabular}
}
\section{Introduction}
The influence of chaotic motion on the dynamics of systems
with a large number of relevant degrees of freedom is one of the central
and unsolved issues of nonlinear dynamics. This problem has a
strong impact on applications (cf.~\cite{CrHo}). 
It is not very well understood from the analytical point of view,
in despite of the enormous amount of numerical results, which are available 
in the literature.
This drawback comes to a certain extent from the lack of simple nontrivial
models, which can be studied by elementary methods, and 
which allow for an investigation of such basic problems. 
The present publication intends to narrow this gap.

Recalling the tremendous success, that low--dimensional time discrete
dynamical systems had played in the understanding of low--dimensional chaos,
systems of coupled maps have become a paradigm for the investigation of
complex space time behaviour \cite{Kanebuch}. 
They allow for both efficient numerical simulations \cite{Kane} and
for analytical approaches. Especially it is possible to introduce a definite 
notion of space time chaos\footnote{As already pointed out by the
authors, the terminology 
space time mixing is more appropriate.} \cite{BuSi,BrKu}.

These analytical and even rigorous approaches are based on what is called
the thermodynamic approach towards dynamical systems. Its main idea
is based on the fact, that owing to a symbolic dynamics
the stationary dynamical properties can be described in terms of the 
statistical mechanics of spin systems. Within such an approach
one dimension of the spin lattice corresponds to the time axis of the 
dynamical system, whereas the other dimensions correspond to the spatial 
extension. Bifurcations in the dynamical system are reflected by
phase transitions in the corresponding spin system. 

Such concepts
are especially easy and elementary to apply within the context
of one--dimensional piecewise linear Markov maps \cite{mori,wang,just93}.
Of course, since the dynamical system has no spatial extension
the associated spin system is one--dimensional, and
long range interactions are necessary to cause a phase transition.
Such interactions are typically related to
a nontrivial grammar, singular expansion rates or the violation of
transitivity in the dynamical system. 

Among
spatially extended systems phase transitions may occur under much
milder conditions since the associated spin lattice is at least 
two--dimensional. In fact, it is already a nontrivial task to establish 
a conventional
high temperature phase, since the interactions which occur in the spin
system are rather intricate. Nevertheless,
it has been shown rigorously that certain weakly 
coupled maps admit correlations decaying exponentially in space and
time, and that the corresponding spin system is in the paramagnetic
phase. Hence it was suggested \cite{buni95} that the occurrence of more
complicated coherent patterns, usually observed in numerical simulations, are
related to the phase transitions in the associated spin system.

There are several examples available in the literature reporting
phase transitions, and I will mention a few of them. For a certain class
of coupled logistic mappings, where a space time mixing state in the
weak coupling regime was proven to exist \cite{Vole}, stable periodic
patterns emerge beyond such a regime \cite{BuCa}. Although these properties
show rigorously that a phase transition occurs, nothing is known about
its nature. Furthermore, the scenario resembles bifurcations in
low--dimensional dynamical systems, since the spatial extension seems to
play no crucial role for the transition. The same conclusion holds for the
so called ''peak crossing bifurcation'' \cite{BuVe}, a kind of boundary
crisis which causes space time intermittency from a trivial non chaotic state.
Finally the model proposed by Miller and Huse \cite{MiHu} shows an Ising
like transition, but only in two spatial dimensions. The model is quite
different from that proposed here, since the local
dynamics used by Miller and Huse mimics to a certain extent a single 
spin. In fact one--dimensional coupled map lattices of this type do 
not indicate a phase transition, but the real mechanism does not
seem to be fully understood \cite{BoBu}.

Phase transitions associated with short range
interactions in the symbolic dynamics should display a finite size behaviour,
since the transition is only possible in the thermodynamic limit, 
i.e.~in the limit of large system size. Hence the corresponding instability
is expected to be related to a transient behaviour, where the length
of the transients increase with the system size. Numerical
simulations show such a behaviour (cf.~\cite{kane90,just95}).
Altogether these considerations indicate that it might be helpful 
to develop a model
which can be handled analytically with moderate effort. 

I will propose a model which possesses a coupling quite different from the
famous ''diffusive'' type coupling\footnote{As already stressed 
in \cite{fuji,buni95} this coupling does not mimic the real diffusion 
but is merely chosen as a simple and typical short range coupling.}.
It is not surprising that a model which can be handled analytically
has a special type of
spatial interaction too. For that reason section \ref{sec2}
is devoted to the introduction of the basic idea using two
coupled maps only. We are lead to the discussion of simplicial mappings 
which are the natural generalisation of piecewise linear one--dimensional
Markov maps. The construction will then be performed for an arbitrary
number of maps in section \ref{sec3}. 
Although we will finally apply the results
for a globally coupled system, the approach is certainly not
restricted to that case as stated in the conclusion. 
\section{Two--dimensional simplicial mapping}\label{sec2}
Recalling that piecewise linear one--dimensional Markov maps can be handled
analytically, it is tempting to construct a coupled map lattice with similar
properties. Particularly,
I want to present a model which is by construction
piecewise linear on its Markov partition, but allows for instabilities,
that means phase transitions, due to the spatial extension. However, before
I dwell on this problem let me introduce in this section
the basic ideas of this construction.  For that purpose
we restrict ourselves to the simple case of two coupled maps only.

\begin{figure}
\begin{center} \mbox{\epsffile{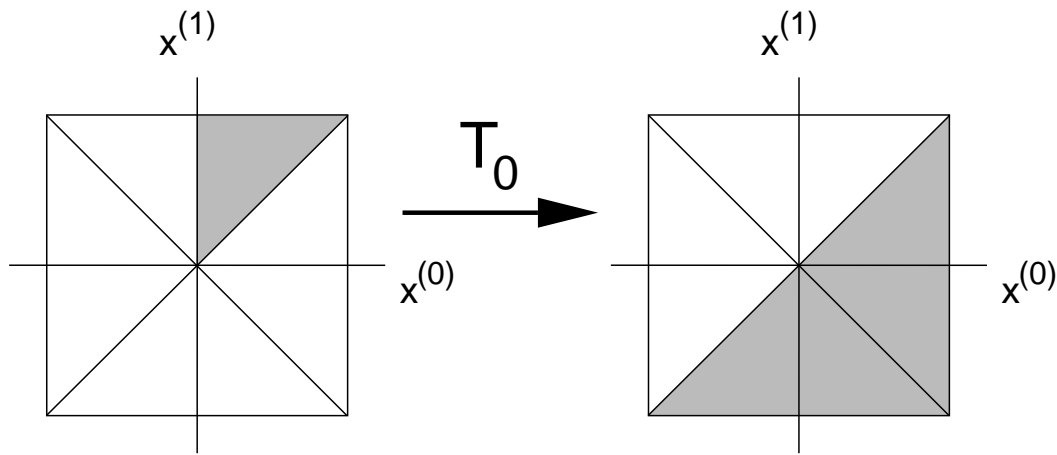}} \end{center}
\caption[ ]{Transformation properties of the uncoupled tent maps (\ref{s2a}). 
A simplex and its image is  indicated.\label{fig_s2a}}
\end{figure}

To begin with let us consider a system of two independent tent maps
\be{s2a}
x_{n+1}^{(\nu)} = \left(T_0 (\ul{x})\right)^{(\nu)}:=
1 - 2 |x_n^{(\nu)}| \quad .
\ee
Here the spatial index $\nu$ takes the values $0$ or $1$. Of course the 
Markov partition of the single map carries over as a direct product
\be{s2b}
I_{\sigma^{(0)} \sigma^{(1)}} := \{ (x^{(0)}, x^{(1)}) \, | \,
0 \leq \sigma^{(\nu)} x^{(\nu)} \leq 1 \},
\quad \sigma^{(\nu)}\in \{-1,1\}  \quad .
\ee
Each set (\ref{s2b}) of the Markov partition is mapped to the full phase space
by the dynamics (cf.~fig.\ref{fig_s2a}).
The grammar is trivial in the sense that all symbol
sequences consisting of pairs $(\sigma^{(0)},\sigma^{(1)})$
are permitted by the dynamics.

\begin{figure}
\begin{center} \mbox{\epsffile{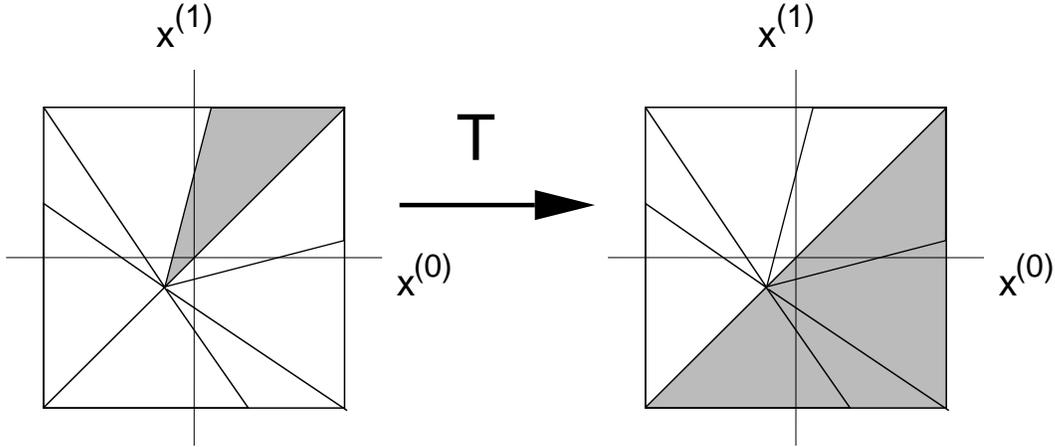}} \end{center}
\caption{Simplices of the Markov partition of two coupled maps obtained
by a deformation of the uncoupled case. The shaded area indicates a
simplex and its image.\label{fig_s2b}}
\end{figure}

We are now going to introduce an interaction in such a way that the map
remains piecewise linear on a suitable Markov partition and hence will
allow for an analytical and elementary
treatment. The strategy consists in fixing a
suitable Markov partition, and then constructing the corresponding
analytical form of the mapping. We start from the Markov
partition of the uncoupled system 
and deform the partition appropriately (cf.~fig.\ref{fig_s2b}). 
Consider a fixed set of the partition we intend to
construct. On this set the two--dimensional map has to be affine. A general 
two--dimensional affine mapping has six adjustable parameters.
Therefore by fixing three points in phase space and their images the
affine mapping is determined uniquely. 
For that reason we cannot stick to the cubes\footnote{If one demands that
the map is continuous and piecewise linear on rectangles, then we are left 
with the uncoupled case.}, which constitute a
Markov partition of the uncoupled system, but have to trace back to
triangles, i.e.~two--dimensional simplices.
In fact, specifying a simplex and an image simplex fixes
uniquely an affine transformation, which maps {  the former set onto the
latter one.}
Therefore one introduces an interaction between the two tent maps, if
one deforms the simplices in the Markov partition but requires that the
images of these simplices coincide with the image simplices
of the uncoupled case (cf.~fig.\ref{fig_s2b}). 
However the deformation cannot be performed arbitrarily, since the image of
each simplex has to consist of a union of simplices in order to guarantee
the Markov property. Hence vertices on the boundary of phase space 
have to stay on the boundary, and the vertex on the diagonal has to stay on
the diagonal. This simple geometrical prescription constitutes the basic idea
for the construction of the coupled map lattice. In what follows a suitable
notation will be introduced, in order to present the analytical expressions
for the map and the associated partition function. The approach will be held
general enough to capture the case of arbitrary dimension too.

To label the simplices again we start from the uncoupled case. The simplices 
are obtained by dividing each square (\ref{s2b}) into two simplices as
indicated in fig.\ref{fig_s2a}. Each simplex has three
vertices, the first being given by a ''corner'' of the phase space 
$(\sigma^{(0)},\sigma^{(1)})$, the second being given by the intersection
of the boundary and the coordinate axis, and the last one is located at
the centre $(0,0)$. Corresponding to these vertices the following 
symbol is assigned to the two resulting simplices
\bea{s2c}
\{ (\sigma^{(0)},\sigma^{(1)}), (0,\sigma^{(1)}), (0,0)\}
&\leftrightarrow&  
[(\sigma^{(0)},\sigma^{(1)}); I] \nonumber\\
\{ (\sigma^{(0)},\sigma^{(1)}),(\sigma^{(0)},0), (0,0) \}
&\leftrightarrow&  
[(\sigma^{(0)},\sigma^{(1)}); Q ] \quad .
\eea
Alternatively this convention corresponds to the notation
\be{s2d}
[(\sigma^{(0)},\sigma^{(1)}); P ] :=
\{ (x^{(0)},x^{(1)})\, | \, 0 \leq \sigma^{P(0)} x^{P(0)} \leq
\sigma^{P(1)} x^{P(1)} \leq 1 \} \quad ,
\ee
if {  $P$ denotes a permutation which is either $Q$, the transposition of the two 
numbers $0,1$ or $I$,} the identity permutation. 
Within this notation the uncoupled map $T_0$ induces the
following transitions, i.e.~the grammar of the system
\bea{s2e}
[ (\sigma^{(0)},\sigma^{(1)});I]
&\rightarrow& [(1,1);Q] \cup [(1,-1);Q] \cup
[ (1,-1);I] \cup [(-1,-1);I] \nonumber\\
\,
[(\sigma^{(0)},\sigma^{(1)});Q]
&\rightarrow& [(1,1);I] \cup [(-1,1);Q] \cup
[ (-1,1);I] \cup [(-1,-1);Q] \quad .
\eea

As already mentioned the interaction between the tent maps is introduced,
by shifting the vertices $(\pm 1,0)$ and $(0, \pm 1)$ along the
boundary an amount, which we denote by $t_{(\pm 1,0)}$ and
$t_{(0,\pm 1)}$ respectively. The shifted vertices read
$(\pm 1, t_{(\pm 1, 0)})$ and $(t_{(0,\pm 1)}, \pm 1)$. In the same way the
vertex on the diagonal $(0,0)$ is shifted to $(t_{(0,0)}, t_{(0,0)})$.
The deformed simplex $[(\sigma^{(0)},\sigma^{(1)});P]$, 
to which we assign the same label, possesses
the vertices
\be{s2f}
[(\sigma^{(0)},\sigma^{(1)}), P] \leftrightarrow
\left\{ \begin{array}{ll}
\{(\sigma^{(0)},\sigma^{(1)}) , (t_{(0,\sigma^{(1)})},\sigma^{(1)}), 
(t_{(0,0)},t_{(0,0)}) \} & P=I \\
\{ (\sigma^{(0)},\sigma^{(1)}) , (\sigma^{(0)}, t_{(\sigma^{(0)},0)}), 
(t_{(0,0)},t_{(0,0)}) \} & P=Q \end{array} \right. \quad .
\ee
The grammar (\ref{s2e}) and the condition that the map acts linearly on
the simplices, define the dynamical system completely. 
Within this setting the numbers
$t_{(\sigma,\sigma')}$ act as parameters for the map $T$.

It is now possible to evaluate the partition function
corresponding to the topological pressure. For that purpose we need the
local expansion rate, i.e.~the Jacobian. The latter
is easily obtained
from the ratio of the volume of a simplex and the volume of its image.
Hence there is no need to write down the closed analytical formula 
for the coupled system at the moment. 
If one introduces the notation $\ul{\sigma}=(\sigma^{(0)},\sigma^{(1)})$
and
\be{s2g}
\ul{\sigma}^{\{0\}} :=(0,\sigma^{(1)}), \quad
\ul{\sigma}^{\{1\}} :=(\sigma^{(0)},0), \quad
\ul{\sigma}^{\{0,1\}} := (0,0)
\ee
then the volume of the simplex (\ref{s2f}) is given by
\bea{s2h}
\mbox{Vol}[(\sigma^{(0)},\sigma^{(1)}), P] &=& \frac{1}{2}
\left| \sigma^{P(0)} - t_{\ul{\sigma}^{P\{0\}}} \right|
\left| \sigma^{P(1)} - t_{\ul{\sigma}^{P\{0,1\}}}\right|\nonumber\\
&=& \frac{1}{2}
\left(1- \sigma^{P(0)} t_{\ul{\sigma}^{P\{0\}}}\right)
\left(1- \sigma^{P(1)} t_{\ul{\sigma}^{P\{0,1\}}}\right) \quad .
\eea
Since the volume of the image
is always given by $2^2/2$ (cf.~fig.\ref{fig_s2b}) one obtains
for the Jacobian the expression
\be{s2i}
\left| \det DT \right|_{[\ul{\sigma},P]} =
2^2 \left( 1 - \sigma^{P(0)} t_{\ul{\sigma}^{P \{0 \}}}\right)^{-1}
\left( 1 - \sigma^{P(1)} t_{\ul{\sigma}^{P\{0,1\}}}\right)^{-1} \quad .
\ee
Performing the summation over all 
$n$--periodic symbol  sequences\footnote{In fact, a slight 
ambiguity arises when
introducing the partition function, since the correspondence
between symbol sequences and orbits is not unique, if orbits on the boundaries
of the partition are considered.}
$[\ul{\sigma}_0;P_0],\ldots,$ $[\ul{\sigma}_{n-1};P_{n-1}],
[\ul{\sigma}_0;P_0],\ldots$ that are permitted by the grammar (\ref{s2e})
yields for the partition sum corresponding to the local expansion rate 
the expression
\be{s2j}
Z_n(\beta) = \sum' \prod_{k=0}^{n-1} 2^{-2 \beta}
\left( 1 - \sigma_k^{P_k(0)} 
t_{\ul{\sigma}_k^{P_k \{0\}}}\right)^{\beta}
\left( 1 - \sigma_k^{P_k(1)} 
t_{\ul{\sigma}_k^{P_k \{0,1\}}}\right)^{\beta} \quad .
\ee
The summation in eq.(\ref{s2j}) involves both a summation over
symbol patterns $\{\sigma_k^{(\nu)}\}$ and over sequences of permutations
$P_k$ which are related by the grammar. In order to read off the
spin Hamiltonian from eq.(\ref{s2j}) the summation with respect to the
permutation sequences has to be performed.
But as indicated in appendix \ref{apb} this 
sequence is almost uniquely determined by the symbol pattern, if surface
terms are discarded. Hence expression (\ref{s2j}) may be understood as a
sum over all periodic symbol patterns $\{\sigma_k^{(\nu)}\}$, with the
sequence of permutations being determined by this pattern 
according to the grammar (\ref{s2e}). Summarising, eq.(\ref{s2j}) represents
the partition sum of two coupled spin chains with the interaction being 
mediated by the deformation parameters $t$, 
i.e.~by the coupling in the dynamical 
system. No phase transition occurs of
course, since no long range interactions are involved. 
\section{Ising type transition in a globally coupled model}\label{sec3}
The considerations of the preceding section will be carried over to the
case of $L$ coupled maps, where $L$ may be understood as 
the extension in the ''spatial''
direction of the dynamical system. 

We start again from $L$ independent tent
maps. Let $\ul{x}=(x^{(0)},\ldots,x^{(L-1)})$ specify the state of the
system in the phase space $[-1,1]^L$. Each cube of the Markov partition, as
usual labelled by $\ul{\sigma}\in\{-1,1\}^L$, is
divided into $L!$ simplices according to (cf.~eq.(\ref{s2d}))
\be{s3a}
[\ul{\sigma};P] := \{ \ul{x}\, | \, 0\leq \sigma^{P(0)} x^{P(0)}\leq
\ldots \leq \sigma^{P(L-1)} x^{P(L-1)}\leq 1 \} \quad ,
\ee
where $P$ denotes a permutation of the
$L$ numbers $0,\ldots,L-1$. The vertices of this simplex are given by
\be{s3b}
\ul{\sigma}, \ul{\sigma}^{P\{0\}},\ldots ,
\ul{\sigma}^{P\{0,\ldots,L-1\}} \quad .
\ee
Here $\ul{\sigma}^{P\{0,\ldots,\nu\}}$ denotes the phase space point
which is obtained by replacing the entries at positions $P(0),\ldots,P(\nu)$ 
in the symbol sequence $\ul{\sigma}$ with zero. Uncoupled tent maps have
the property, that the image $T_0(\ul{\sigma}^{P\{0,\ldots,\nu\}})$ of a
vertex is obtained, if each coordinate with value $\pm 1$ is replaced
by $-1$, and each coordinate with value $0$ is replaced by $+1$.
Using this rule the grammar corresponding to the partition (\ref{s3a}) 
can be constructed, but we will not need these expressions in what
follows.

The interaction among the maps is introduced by deforming the simplices,
i.~e.~by shifting the coordinates of the vertices (\ref{s3b}) which are zero
by a constant amount. The shifted vertex is generated from 
$\ul{\sigma}^{P\{0,\ldots,\nu\}}$ by replacing every zero in the sequence
by the same constant value $t_{\ul{\sigma}^{P\{0,\ldots,\nu\}}}$, 
which of course is smaller than $1$ in modulus
(cf.~eqs.(\ref{s2c}) and (\ref{s2f})). 
The shift may differ
from vertex to vertex as indicated by the subscript. For the
shifted vertex the notation  $\ul{\tau}_{\ul{\sigma}^{P\{0,\ldots,\nu\}}}$
is introduced.
The construction of the coupled map lattice is completed by the condition
that the image of the vertex is given by the image of the corresponding
vertex of the uncoupled system
$T(\ul{\tau}_{\ul{\sigma}^{P\{0,\ldots,\nu\}}})=
T_0(\ul{\sigma}^{P\{0,\ldots,\nu\}} )$. 
{  The last condition guarantees that the image of each simplex is
given by a union of simplices like in the uncoupled case.}
Since a closed analytical formula
of the coupled map lattice $T$ will be of no use for us, I refrain from
writing down such an expression. However for numerical purposes 
appendix \ref{apc} contains a simple recursive algorithm to evaluate
$T$. 

In order to determine the Jacobian one needs the volume of the deformed 
simplices. The latter follows immediately from the vertices described above as
\be{s3c}
\mbox{Vol}[\ul{\sigma};P] = \frac{1}{L!} \prod_{\nu=0}^{L-1}
\left( 1 - \sigma^{P(\nu)} t_{\ul{\sigma}^{P\{0,\ldots,\nu\}}} \right) \quad .
\ee
Since the image of each simplex has the volume $2^L/L!$ the Jacobian reads
\be{s3d}
\left| \det DT \right|_{[\ul{\sigma};P]} =
2^L \prod_{\nu=0}^{L-1}  
\left( 1 - \sigma^{P(\nu)} t_{\ul{\sigma}^{P\{0,\ldots,\nu\}}} \right)^{-1}
\quad .
\ee
Hence we obtain for the partition sum 
\be{s3e}
Z_n(\beta) = \sum' \prod_{k=0}^{n-1} 2^{- L \beta}
\prod_{\nu=0}^{L-1}
\left( 1 - \sigma_k^{P_k(\nu)} 
t_{\ul{\sigma}_k^{P_k \{0,\ldots,\nu\}}}\right)^{\beta} \quad ,
\ee
where again the summation has to be performed with respect to
all $n$--periodic accessible symbol sequences 
$[\ul{\sigma}_0;P_0],\ldots,[\ul{\sigma}_{n-1};P_0],
[\ul{\sigma}_0;P_0],\ldots$. As already indicated in the preceding
section the permutation sequence $P_k$ is (almost) uniquely determined
by the two--dimensional symbol pattern $\{ \sigma_k^{(\nu)}\}$.
Hence we may alternatively
perform the summation with respect to all the spin variables,
with the permutations being determined by the spin pattern.
Then the corresponding Hamiltonian of the two--dimensional spin
lattice can be read off from eq.(\ref{s3e}) directly.

{  The structure of the Hamiltonian is however a little bit hidden
since the permutations depend on the spin variables.
To simplify the subsequent considerations we will refer in what follows
to a special choice which corresponds to a globally coupled system, so that
the permutations drop from the formulas.
Let us choose the deformations in such a way, that on the one hand
no interaction in the index $k$, i.e.~in the direction of time occurs, and
on the other hand a global interaction between the spins in the 
index $\nu$, i.e.~an interaction in the spatial direction 
is realised. Then we are left with simple mean field Ising chains.}
\be{s3f}
\prod_{\nu=0}^{L-1}
\left( 1 - \sigma^{P(\nu)} 
t_{\ul{\sigma}^{P \{0,\ldots,\nu\}}}\right)^{\beta}=
\exp\left( \beta H \sum_{\nu=0}^{L-1} \sigma^{(\nu)} + \beta \frac{J}{L}
\sum_{\nu,\mu =0}^{L-1} \sigma^{(\nu)} \sigma^{(\mu)}+ \beta E_0 \right)
\ee
Confer appendix \ref{apd} for the corresponding explicit 
construction, which determines the deformation parameters 
$t_{\ul{\sigma}^{P\{0,\ldots,\nu\}}}$ in terms of $H$
and $J$ and the normalisation constant $E_0$. 
Roughly speaking the parameter $H$, corresponding to the magnetic field
in the spin system, describes the deviation of the map lattice from an 
inversion symmetric case, whereas the spin interaction $J$ mediates the 
spatial coupling among the maps. Inspecting eqs.(\ref{s3e}) and (\ref{s3f}),
the Hamiltonian of the full map lattice consists of non interacting 
spin chains. Since the interaction within a chain has infinite range 
a phase transition occurs at $\beta H=0$, 
$\beta J=1/2$ if the thermodynamic limit $L\rightarrow \infty$
has been performed. Hence our coupled map lattice displays a bifurcation,
which is intimately related to the limit of large system size. Of course
the model allows for an analytical calculation of those quantities,
which are determined by the spin pattern, since the partition sum can
be evaluated in the thermodynamic limit.

\begin{figure}
\begin{center} \mbox{\epsffile{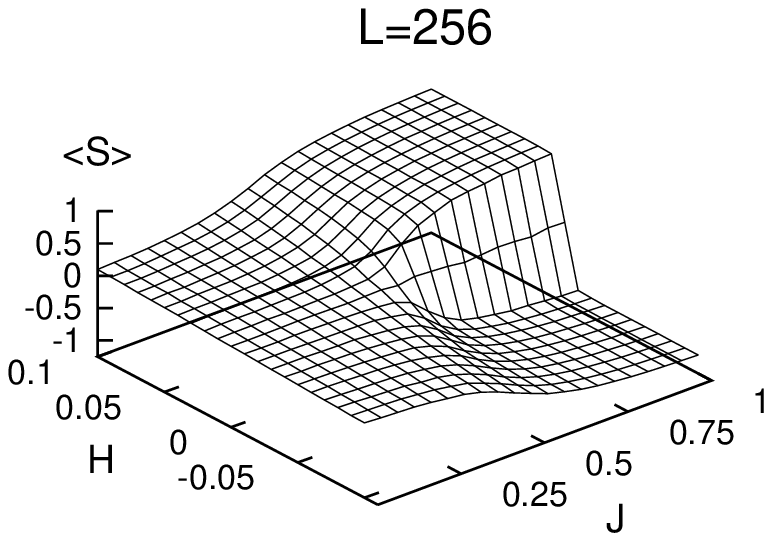}}\\[5mm]
\mbox{\epsffile{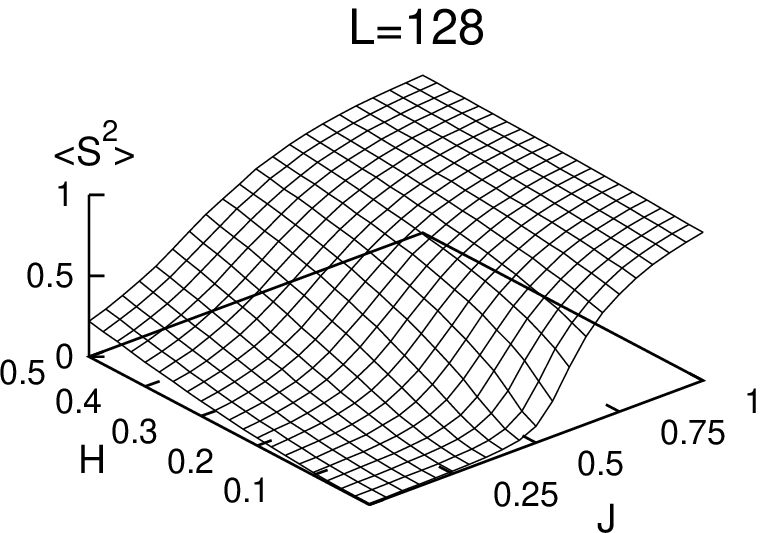}}
\end{center}
\caption{Dependence of the spatially averaged symbol and its square
on $H$ and $J$ for system size $L=256$ and $L=128$. The time average has been
performed using a spatial uniformly distributed initial condition and $2000$ 
iteration steps, after discarding a transient of length $200$.\label{fig_s3a}}
\end{figure}

Let us supplement the analytical results with numerical
simulations of the full map lattice (cf.~appendix \ref{apc} and \ref{apd}
for the numerical algorithm). Keep in mind however that the usual long time
average, which will be denoted by $\langle \ldots \rangle$ in the sequel
corresponds to the choice $\beta=1$ in the context of dynamical systems.
Obvious global quantities are given by the spatially averaged symbol
sequence and its square
\be{s3g}
\langle S\rangle :=\left\langle \frac{1}{L} \sum_{\nu=0}^{L-1}
\sigma^{(\nu)} \right\rangle, \qquad
\langle S^2 \rangle :=\left\langle \left( \frac{1}{L} \sum_{\nu=0}^{L-1}
\sigma^{(\nu)} \right)^2 \right\rangle \quad .
\ee
The dependence of these averages on $H$ and $J$ is displayed in
fig.\ref{fig_s3a}. The well known behaviour of the Ising mean field
model is reproduced, but of course 
the results have been obtained here from the dynamics of the map lattice.
{  One should however keep in mind
that from the point of view of the dynamical system the quantities 
(\ref{s3g}) look
a little bit artificial and the actual interpretation of the dependence
on the system parameters is difficult. Nevertheless the main feature,
i.~e.~the} second order phase transition clearly shows up at $H=0$, $J=1/2$,
but finite size effects are superimposed. 

\begin{figure}
\begin{center} \mbox{\epsffile{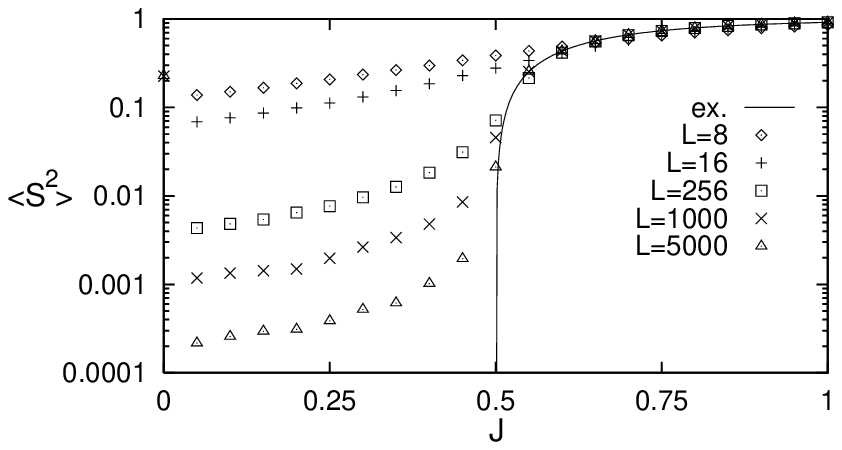}} \end{center}
\caption{Dependence of $\langle S^2 \rangle$ on the interaction $J$ for
zero field $H=0$ and different system sizes (symbols). The solid line
displays the analytical result in the thermodynamic limit 
$L\rightarrow \infty$.\label{fig_s3b}}
\end{figure}

In order to demonstrate the 
influence of the system size on the transition more clearly, consider the
zero field dependence of $\langle S^2\rangle$ on the interaction strength
for different system size (fig.\ref{fig_s3b}).
Since the phase transition sharpens if the thermodynamic limit is
approached, it is obvious that the transition in the dynamical system
is intimately related to the thermodynamic limit. Hence it is a
characteristic feature of the high--dimensional dynamical system.

\begin{figure}
\begin{center} \mbox{\epsffile{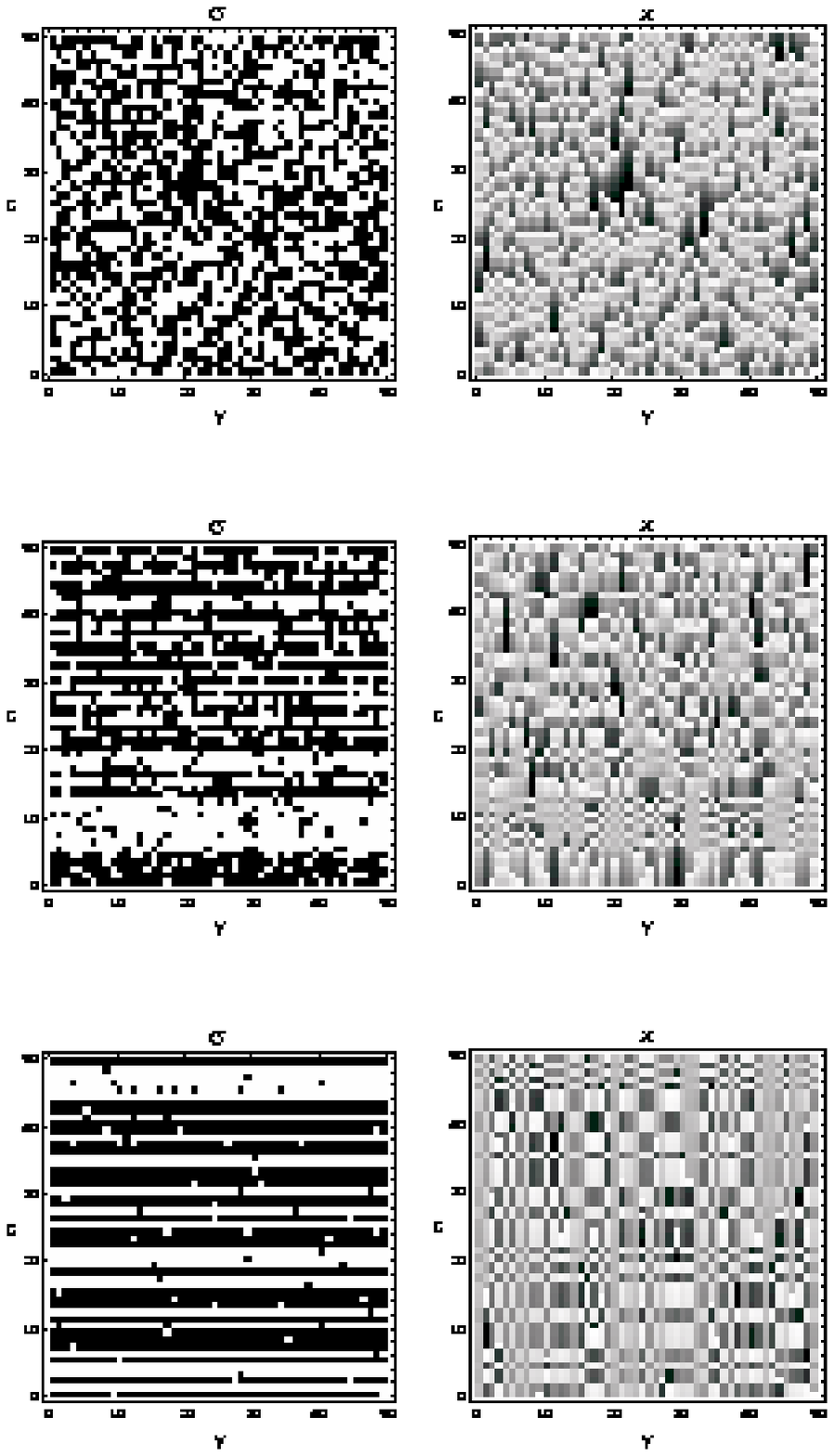}} \end{center}
\caption{Space time evolution of the symbols $\sigma_k^{(\nu)}$ and
of the phase space coordinate $x_k^{(\nu)}$ for a system of size $L=256$,
field $H=0.0$, and coupling strengths increasing from above 
to below $J=0.2,0.6,1.0$. 
Black/white pixels correspond to the values $\sigma=-1/+1$ and dark/light
pixels to phase space coordinates close to $x=-1/+1$.\label{fig_s3c}}
\end{figure}

\begin{figure}
\begin{center} \mbox{\epsffile{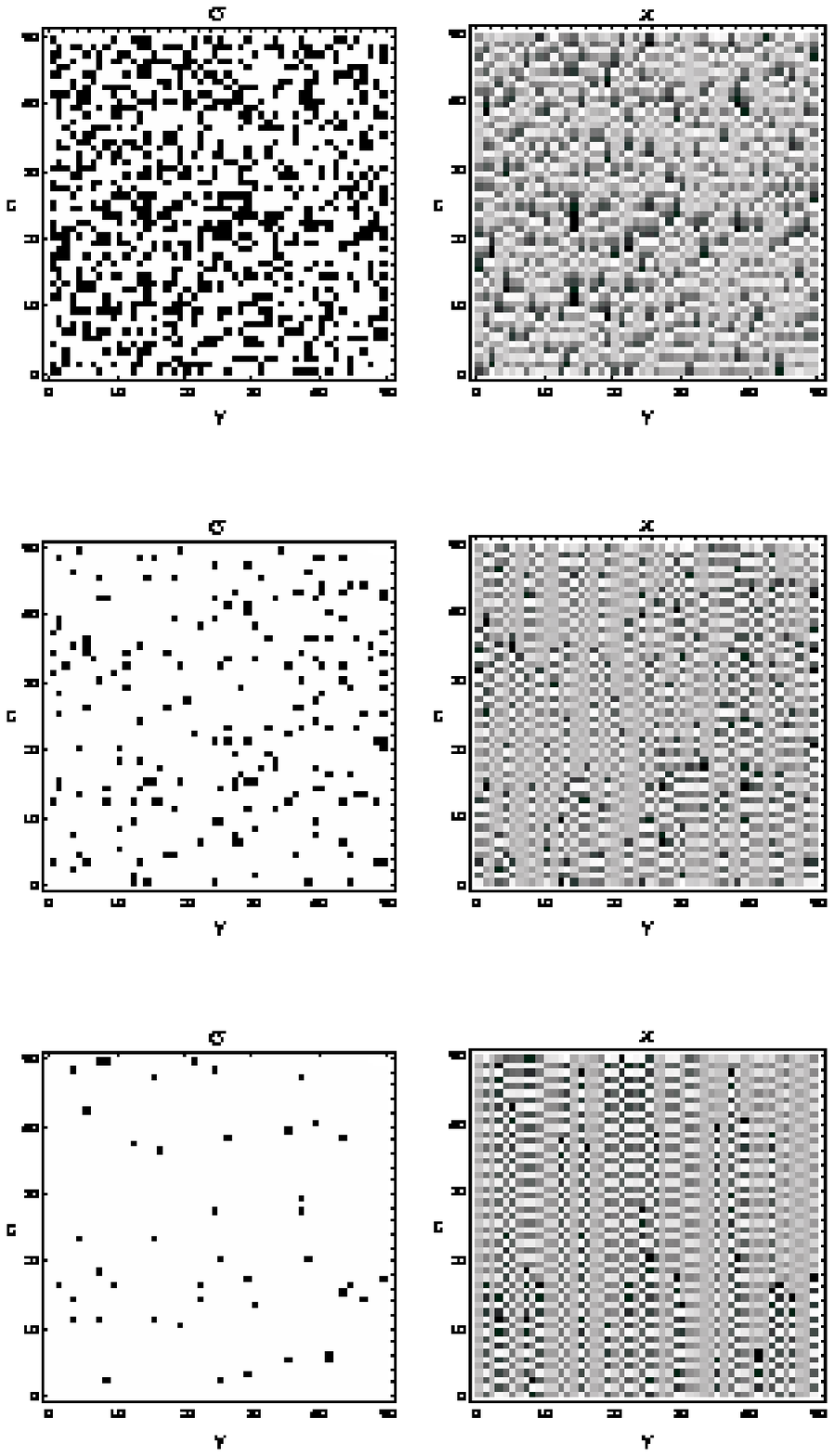}} \end{center}
\caption{Space time evolution of the symbols $\sigma_k^{(\nu)}$ and
of the phase space coordinate $x_k^{(\nu)}$ for a system of size $L=256$,
field $H=0.2$, and coupling strengths increasing from above 
to below $J=0.2,0.6,1.0$. 
Black/white pixels correspond to the values $\sigma=-1/+1$ and dark/light
pixels to phase space coordinates close to $x=-1/+1$.\label{fig_s3d}}
\end{figure}

Finally, let us have a look at the space time evolution in 
symbol space as well as in phase space. Consider first the evolution
in an inversion symmetric situation, i.e.~$H=0.0$ (cf.~fig.~\ref{fig_s3c}).
Below the transition point, i.~e.~for weak coupling, a rather irregular
state is observed. It corresponds to the space time chaotic regime mentioned
in the introduction. Slightly above the transition point long range order
develops, but a lot of ''defects'' are superimposed. The synchronisation
becomes more perfect, if the coupling is further strengthened. Of course both
phases are mixed, since no symmetry breaking field is present.
With a symmetry breaking field the weak coupling regime behaves very similar
to the zero field case (c.f.~fig.\ref{fig_s3d}).
But above the transition point one phase is dominant and a pure state
develops if the coupling is increased. In summary,
the phase transition shows up in the symbol representation as well as 
in the phase space coordinates, where apparently two different kinds
of motion occur.
\section{Conclusion}
A class of piecewise linear coupled map lattices has been investigated
which allows for an analytical approach. The key point of the construction
consists in fixing a Markov partition in terms of simplices, imposing
a grammar, and requiring
the map to be linear on this partition. Such an approach allows for an 
elementary and analytical construction of the corresponding spin Hamiltonian.
Phase transitions are possible at finite coupling in the limit of large 
system size, although the model is locally expanding and has the Markov 
property. 

The proposed construction may be understood as the natural
extension of piecewise linear one--dimensional 
maps to more than one dimension. 
In fact, by a refinement of the partition a large class of functions, at least
in finite dimensions, can be  approximated, as stated by the simplicial 
approximation theorem. Furthermore, piecewise linear one--dimensional maps are 
able to approximate
dynamical features of low--dimensional chaos, as stated by the 
Ulam conjecture \cite{BlKe}. Similar properties might hold also for simplicial
mappings in connection with high--dimensional dynamics, 
but further clarification is certainly needed.

The explicit evaluation of a phase transition has been performed for a 
global coupling only. Although globally coupled systems are interesting in 
itself I have dealt with a globally coupled system for reasons of 
simplicity, since otherwise the exact computation of the
phase transition may become much more involved. 
I believe that the principal aspects of this 
paper do not depend on this choice of coupling.
Indeed an interaction in the time direction of the spin 
system may be introduced {  by tracing back to finer partitions}, 
and a spatial short range coupling may be realised.
Hence it seems possible to construct a model along these lines 
{  which is in some sense closer to a nearest neighbour coupled
Ising model in two dimensions, i.~e.~a model} with
short range interactions in each direction but admitting a phase transition
in the thermodynamic limit. Since the objective of
the present publication was on the principal aspect I leave this problem
for future work. 

{  In the treatment I have adopted a physicists approach, by
considering systems of arbitrary but finite size and performing the
thermodynamic limit at the end of the calculation. Therefore it is not
easy to decide rigorously whether the dynamical system, being defined 
by a geometrical construction, tends to a mathematically well defined
limit. Nevertheless the discussion of the example supports such a conjecture.
Then the phase transition like behaviour in the finite model should be
related to a transient dynamics whose duration increases with the system size.
Such an analysis, which is of course desirable for a deeper
understanding of the transition has not been accomplished yet.}

Models of the type investigated here, i.~e.~models with a large number
of degrees of freedom showing a phase transition and allowing for an
elementary analytical approach, may be helpful to understand several features
of high--dimensional chaotic dynamics. Although 
the grammar, i.~e.~the structure
of space time periodic orbits was fixed by hand, it is also possible to
investigate the influence of pruning, by considering increasingly finer
Markov partitions. These properties together with an analytical accessible
system may have an impact on the understanding of expansions in 
terms of space time periodic orbits, which have recently attracted
some interest.
\section*{Acknowledgement}
I am grateful to F.~Christiansen for critical reading of the manuscript.
A part of this work was performed at the TH--Darmstadt within a program
of the Sonderforschungsbereich 185 ''Nichtlineare Dynamik''.
\appendix
\section{Two coupled maps}\label{apb}
In order to simplify our computations and not hide the main arguments
behind extensive computations let us specialise to the case
$t_{(0,\sigma )}=t_{(\sigma,0)}= - \sigma \tanh J $, $t_{(0,0)}=0$.
I think it will become
evident that this choice does not influence the generality of the arguments.

The partition sum (\ref{s2j}) is taken with respect to all $n$--periodic
sequences $([\ul{\sigma}_k ;P_k])$. Consider first those patterns
which obey $\sigma_k^{(0)} \neq \sigma_k^{(1)}$
for some index $k\in [0,n-1]$. Since the symbols $(\sigma^{(0)},\sigma^{(1)})$,
that means the sets $[\ul{\sigma};I]\cup[\ul{\sigma};Q]$,
already determine a Markov partition, the pattern $\{\sigma_k^{(\nu)} \}$ 
specifies a phase space point, which in our case is a period--$n$ 
orbit\footnote{{  For the particular case
$\sigma_k^{(0)}\equiv 1$, $\sigma_k^{(1)}\equiv -1$ such an argument
does not apply. But this single point does not matter in the
thermodynamic limit.}}. 
But this orbit does not hit a boundary of the simplices and especially
not the diagonal of the phase space, since at least one
pair of symbols has different entries. The validity of this 
statement is obvious
for the case of noninteracting maps. It carries over to the general case, 
since the deformation of the simplices does not destroy topological
properties in the phase space. Hence the sequence of 
permutations $P_k^{(\nu)}$ is determined uniquely by the periodic orbit, that
means by the pattern of symbols $\{\sigma_k^{(\nu)} \}$. 

Consider instead $\sigma_k^{(0)} = \sigma_k^{(1)}$ for all
$k$, i.e.~a synchronised orbit. Then the corresponding phase space point is
located on the diagonal, i.~e.~on the boundary of the simplices.
From the grammar 
(\ref{s2e}) we have that the predecessor $P_k$ of a permutation $P_{k+1}$
changes according to the symbol $\sigma_{k+1}^{(0/1)}$ and obeys
\be{apba}
P_k = P_{k+1} \mbox{ if } \sigma_{k+1}^{(0/1)}= -1 \qquad
P_k = Q P_{k+1} \mbox{ if } \sigma_{k+1}^{(0/1)}= 1 \quad . 
\ee
The condition that the sequence $P_k$ is $n$--periodic, $P_0=P_n$, requires
that the finite sequence $(\sigma_0^{(0/1)},\ldots,\sigma_{n-1}^{(0/1)})$
contains an even number of symbols $+1$. Furthermore there exist
two corresponding permutation sequences according to eq.(\ref{apba}) 
and the choice $P_0=I,Q$.

Then the partition sum (\ref{s2j}) splits into two parts,
one part containing the summation over non synchronised symbol
patterns, and the second part containing
the summation over synchronised symbol 
patterns with an even number of $+1$ symbols and two complementary
permutation sequences. If we supplement the
first part with the missing synchronised symbol sequences\footnote{In
the following an arbitrary but {  henceforth fixed} 
construction rule for the permutation
sequences is imposed.} then eq.(\ref{s2j}) reads
\bea{apbc}
Z_n(\beta) &=& \sum_{\sigma_k^{(0)}, \sigma_k^{(1)} }
\prod_{k=0}^{n-1} 2^{-2 \beta}
\left( 1 - \sigma_k^{P_k(0)} t_{\ul{\sigma}_k^{P_k
\{0\}}}\right)^{\beta}
\left( 1 - \sigma_k^{P_k(1)} t_{\ul{\sigma}_k^{P_k
\{0,1\}}}\right)^{\beta} \nonumber\\
&+&
\sum_{\sigma_k^{(0)}=\sigma_k^{(1)}, even}
\prod_{k=0}^{n-1} 2^{-2 \beta}
\left( 1 - \sigma_k^{P_k(0)} t_{\ul{\sigma}_k^{P_k
\{0\}}}\right)^{\beta}
\left( 1 - \sigma_k^{P_k(1)} t_{\ul{\sigma}_k^{P_k
\{0,1\}}}\right)^{\beta} \nonumber\\
&-&
\sum_{\sigma_k^{(0)}=\sigma_k^{(1)}, odd}
\prod_{k=0}^{n-1} 2^{-2 \beta}
\left( 1 - \sigma_k^{P_k(0)} t_{\ul{\sigma}_k^{P_k
\{0\}}}\right)^{\beta}
\left( 1 - \sigma_k^{P_k(1)} t_{\ul{\sigma}_k^{P_k
\{0,1\}}}\right)^{\beta} 
\eea
where the index even/odd means, that the summation is performed with
respect to sequences containing an even/odd number of $+1$ symbols.
Now in each sum the sequence of permutations is uniquely
determined by the symbol pattern $\{ \sigma_k^{(\nu)} \}$. 
By our choice of deformations the relation
\be{apbd}
\left( 1 - \sigma_k^{P_k(0)} t_{\ul{\sigma}_k^{P_k
\{0\}}}\right)^{\beta}
\left( 1 - \sigma_k^{P_k(1)} t_{\ul{\sigma}_k^{P_k
\{0,1\}}}\right)^{\beta} 
= \left( 1 + \sigma_k^{(0)} \sigma_k^{(1)} \tanh J \right)^{\beta}
\ee
holds, and it is exactly this property which simplifies the subsequent
considerations. Furthermore we perform in the second and third sum 
of eq.(\ref{apbc}) an
unconstrained summation over $\sigma_k$, $0\leq k \leq n-2$,
choosing $\sigma_{n-1}$ in such a way that the constraint of an even or
odd number of $+1$ values is guaranteed. Then we get
\bea{apbe}
Z_n(\beta) &=& \sum_{\sigma_k^{(0)},\sigma_k^{(1)}}
\prod_{k=0}^{n-1} 2^{-2 \beta}
\left( 1 + \sigma_k^{(0)} \sigma_k^{(1)} \tanh J  
\right)^{\beta} \\
&+&
\sum_{\sigma_0,\ldots,\sigma_{n-2}}
\prod_{k=0}^{n-2} 2^{-2 \beta}
\left( 1 + \tanh J \right)^{\beta} \left\{
\left( 1 + \tanh J \right)^{\beta} -
\left( 1 + \tanh J \right)^{\beta} \right\} \quad .\nonumber
\eea
The second contribution vanishes identically\footnote{In general,
that means for maps which are not symmetric with respect to the origin,
this contribution seems to be a surface term, i.e.~it can be neglected 
compared to the first sum if the exponential $n$--dependence is considered.}.
Now the Hamiltonian can be read off directly from the partition sum
\be{apbf}
-\beta {\cal H} = - \beta \sum_{k=0}^{n-1}\left (
- J \sigma_k^{(0)} \sigma_k^{(1)} + \ln \cosh J + \ln 4 \right) \quad .
\ee
We end up with a simple pair interaction in the ''spatial'' direction and
no interaction in the ''time'' direction.
\section{$L$--dimensional simplicial mapping}\label{apc}
\begin{figure}
\begin{center} \mbox{\epsffile{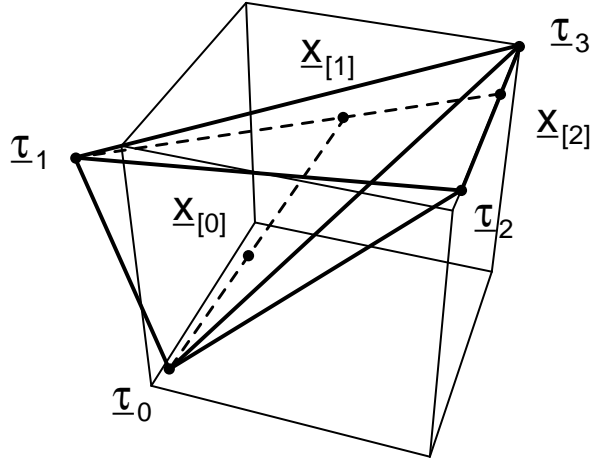}} \end{center}
\caption{Diagrammatic view of the recursive projections
onto lower--dimensional faces of the simplex, which are used to
determine the simplex containing the phase space point $\ul{x}$.
The box indicates a quadrant $[0,1]^3$, $\ul{\tau}_k$ denote the vertices
of the simplex, and $\ul{x}_{[k]}$ the sequence of the projected 
points.\label{fig_apd}}
\end{figure}

For a given set of deformation parameters 
$t_{\ul{\sigma}^{P\{0,\ldots,\nu\}}}$ a
recursive algorithm will be developed, which allows for the calculation
of the image point $T(\ul{x})$. {  Let us assume that $\ul{x}$ is contained
in the interior of a simplex, which from the point of view of a numerical 
evaluation causes no restriction for the continuous mapping $T$.}
Since the simplices are deformed, it is
a nontrivial task to determine the simplex, that contains a given
phase space point $\ul{x}$. However using the fact that simplices are convex, 
a sequence of projections is adopted to determine the corresponding set, 
the simplicial coordinates and hence the image point (cf.~fig.\ref{fig_apd}). 

Let $\ul{x}_{[0]}=
\ul{x}$ denote the phase space point, and $\ul{\tau}_0:=
\ul{\tau}_{\ul{\sigma}^{P\{0,\ldots,L-1\}}}=(t_0,t_0,\ldots,t_0)$
with $t_0:=t_{\ul{\sigma}^{P\{0,\ldots,L-1\}}}=
t_{(0,\ldots,0)}$ the central vertex in the cube $[-1,1]^L$.
By our construction this vertex is common to all the simplices and hence
is a vertex of the simplex which contains the phase space point $\ul{x}$. 
Consider the semi--infinite ray emanating from $\ul{\tau}_0$ through 
$\ul{x}_{[0]}$. It intersects the surface of the cube $[-1,1]^L$ at 
a point denoted by $\ul{x}_{[1]}$
\be{apcz}
\ul{x}_{[1]} = s_0 \ul{x}_{[0]} +(1-s_0) \ul{\tau}_0 \quad .
\ee
Here $s_0$ denotes the number which makes the expression
\be{apca}
s_0:=\frac{1}{|x^{(k_0)}_{[0]}-t_0|}-\frac{t_0}{x^{(k_0)}_{[0]}-t_0}
\rightarrow {\min}
\ee
minimal among all components $k$. The component $k_0$ for which the minimum
is attained determines the surface of the cube where the intersection occurs
by the condition $x^{(k_0)}=\sigma^{(k_0)}$, with $\sigma^{(k_0)}$ 
being given by
\be{apcb}
\sigma^{(k_0)}=\sgn (x^{(k_0)}_{[0]}- t_0)\quad .
\ee
{  That surface contains one face of the simplex we are looking for.}
If we define $P(L-1):=k_0$,
then the central vertex within this surface reads $\ul{\tau}_1:=
\ul{\tau}_{\ul{\sigma}^{P\{0,\ldots,L-2\}}}$, and it leads to the 
second vertex of the simplex we are looking for. 
Now one repeats the same reasoning using the points
$\ul{x}_{[1]}$ and $\ul{\tau}_1$, but with the dimension of the problem
being reduced by one.

Hence one continues by induction. Determine
$k_{\mu}\not\in \{P(L-1),\ldots,P(L-\mu)\}$ among those components, 
which have not been fixed yet, such that
\be{apcd}
s_{\mu}:=\frac{1}{|x^{(k_{\mu})}_{[\mu]}-t_{\mu}|}
-\frac{t_{\mu}}{x^{(k_{\mu})}_{[\mu]}-t_{\mu}}
\rightarrow {\min}
\ee
is minimal. Here $t_{\mu}:=t_{\ul{\sigma}^{P\{0,\ldots,L-1-\mu\}}}$. Then
\be{apce}
\sigma^{(k_{\mu})}:=\sgn (x^{(k_{\mu})}_{[\mu]}- t_{\mu})
\ee
and $P(L-1-\mu):=k_{\mu}$ determine the next components of the symbol
sequence and the permutation respectively.
\be{apcc}
\ul{x}_{[\mu+1]} = s_{\mu} \ul{x}_{[\mu]} +(1-s_{\mu}) \ul{\tau}_{\mu}
\ee
yields the new intersection point, using the abbreviation
$\ul{\tau}_{\mu}=\ul{\tau}_{\ul{\sigma}^{P\{0,\ldots,L-1-\mu\}}}$. 
This iteration terminates exactly
after $L$ steps, since we end up with a corner point of the cube
$[-1,1]^L$. It yields the
simplex $[\ul{\sigma};P]$ which contains the point $\ul{x}$. 

In addition we can read off the simplicial coordinates 
$0\leq \lambda_{\mu} \leq 1$ which are defined by
\be{apcf}
\ul{x}=\sum_{\mu=0}^{L} \lambda_{\mu} \ul{\tau}_{\mu}, \qquad
\sum_{\mu=0}^{L} \lambda_{\mu} = 1 \quad .
\ee
These coordinates are particularly useful for the evaluation of the
image point $T(\ul{x})$. Recalling that the map $T$ is linear on the simplex,
and that the images of the vertices $\ul{\tau}_{\mu}$ are obtained, 
by replacing the components with values $\pm 1 $ by $-1$ and the 
components with values $t_{\mu}$ by $+1$, we get
\bea{apcg}
\left( T (\ul{x})\right)^{P(\nu)} &=& \sum_{\mu=0}^{\nu} \lambda_{\mu}
\left( T (\ul{\tau}_{\mu} )\right)^{P(\nu)} +
\sum_{\mu=\nu+1}^{L} \lambda_{\mu} \left( T (\ul{\tau}_{\mu} )\right)^{P(\nu)}
= \sum_{\mu=0}^{\nu} \lambda_{\mu} -
\sum_{\mu=\nu+1}^{L} \lambda_{\mu}\nonumber \\
&=& -1 + 2 \sum_{\mu=0}^{\nu} \lambda_{\mu} \quad .
\eea
Furthermore we can express the simplicial coordinates in terms of the
quantities (\ref{apcd}), which have already been determined in the course
of the {  recursion. As we prove at the end of this appendix}
\be{apch}
\sum_{\mu=0}^{\nu} \lambda_{\mu} = 1 - \left(\prod_{\mu=0}^{\nu}
s_{\mu}\right)^{-1} 
\ee
{  holds for $0\leq \nu\leq L-1$.}
Hence the simple recursion scheme described above together with 
eqs.(\ref{apcg}) and (\ref{apch}) complete the algorithm for the numerical
evaluation of the map lattice.

We are left with verifying eq.(\ref{apch}). 
Again an induction will be used. First
the component $P(L-1)$ of eq.(\ref{apcf}) yields
\be{apci}
x^{P(L-1)}= \lambda_0 t_0 + (1-\lambda_0)\sigma^{P(L-1)} \quad .
\ee
Then using the definition (\ref{apcb}) one obtains
\be{apcj}
\frac{1}{1-\lambda_0} = \frac{1}{|x_{[0]}^{P(L-1)} - t_0|}
-\frac{t_0}{x_{[0]}^{P(L-1)}-t_0}\quad ,
\ee
which by virtue of eq.(\ref{apca}) leads to the result (\ref{apch}) for
$\nu=0$. Suppose that (\ref{apch}) holds for $\nu\leq \rho$. 
Evaluation of eq.(\ref{apcf}) for the component $P(L-\rho-2)$ yields
\be{apck}
x^{P(L-\rho-2)}_{[0]} = \lambda_0 t_0 + \ldots + \lambda_{\rho+1} t_{\rho+1}
+ ( 1-\lambda_0 - \ldots -\lambda_{\rho+1}) \sigma^{P(L-\rho-2)} \quad .
\ee
On the other hand, we obtain from the recursion relation (\ref{apcc})
\bea{apcl}
x_{[\rho+1]}^{P(L-\rho-2)}-t_{\rho+1} &=& (s_{0} \ldots s_{\rho}) \left(
x_{[0]}^{P(L-\rho-2)} + \frac{(1-s_0) t_0}{s_0} +
\frac{(1-s_1) t_1}{s_0 s_1} + \ldots \right . \nonumber\\
&+& \left. 
\frac{(1-s_{\rho}) t_{\rho}}{s_0 \ldots s_{\rho}} -
\frac{t_{\rho+1}}{s_0 \ldots s_{\rho}} \right) \quad .
\eea
If one now inserts eq.(\ref{apck}) into eq.(\ref{apcl}) and repeatedly uses
eq.(\ref{apch}) for $\nu\leq\rho$ one gets
\bea{apcm}
 & & x_{[\rho+1]}^{P(L-\rho-2)}-t_{\rho+1} \nonumber \\
&=& (s_{0} \ldots s_{\rho}) \left(
(1-\lambda_0-\ldots -\lambda_{\rho+1})\sigma^{P(L-\rho-2)} +
\lambda_{\rho+1} t_{\rho+1} - \frac{t_{\rho+1}}{s_0 \ldots s_{\rho}} \right)
\nonumber\\
&=&
(1-\lambda_0-\ldots -\lambda_{\rho+1}) (s_{0} \ldots s_{\rho}) 
\left( \sigma^{P(L-\rho-2)} -t_{\rho+1} \right) \quad .
\eea
But then from the definition (\ref{apce}) we have
\be{apcn}
\frac{1}{1- \lambda_0 - \ldots \lambda_{\rho+1}} =
(s_{0} \ldots s_{\rho})  \left( \frac{1}{|x_{[\rho+1]}^{P(L-\rho-2)}-
t_{\rho+1}|}-
\frac{t_{\rho+1}}{x_{[\rho+1]}^{P(L-\rho-2)}-t_{\rho+1}} \right)
\ee
and the result follows from the definition (\ref{apcd}).
\section{Spatial mean field coupling}\label{apd}
Using the abbreviation
\be{apda}
S_{\nu}:= \sum_{\mu=\nu+1}^{L-1} \sigma^{(\mu)}
\ee
a sequence of functions $g_{\nu}(S_{\nu})$ will be constructed 
which obeys
\be{apdb}
\sum_{\nu=0}^{L-1} \sigma^{(\nu)}\left( \frac{2 J}{L} S_{\nu} +H \right)
+ J + E_0
= 
\sum_{\nu=0}^{L-1} \ln \left( 1 - \sigma^{(\nu)} \tanh g_{\nu}(S_{\nu})
\right)
\ee
with a constant $E_0$ being independent of the spin variables. Since
the left hand side is invariant with respect to a permutation of the spin
numbers the result (\ref{s3f}) follows then immediately if one defines
\be{apdba}
t_{\ul{\sigma}^{P\{0,\ldots,\nu\}}} :=
\tanh g_{\nu}\left(\sum_{\mu\not\in P\{0,\ldots,\nu\}} \sigma^{(\mu)}\right)
\quad .
\ee

Applying the trivial identity
\be{apdc}
f(\sigma) = \sigma [f(+1)-f(-1)]/2 + [f(+1)+f(-1)]/2
\ee
one obtains
\be{apdd}
\ln \left( 1 - \sigma^{(\nu)} \tanh g_{\nu}(S_{\nu})
\right) = - \sigma^{(\nu)} g_{\nu}(S_{\nu}) - h_{\nu}(S_{\nu})
\ee
if the abbreviation $h_{\nu}(x):= \ln \cosh g_{\nu}(x)$ is introduced.
Observing 
$S_{\nu}=\sigma^{(\nu+1)}+S_{\nu+1}$ and applying eq.(\ref{apdc})
again to the second term on the right hand side one gets
\bea{apde}
 & & h_{\nu}(S_{\nu})  \\
&=& \sigma^{(\nu+1)} \frac{h_{\nu}(1+S_{\nu+1}) -
h_{\nu}(-1+S_{\nu+1})}{2} +
\frac{ h_{\nu}(1+S_{\nu+1}) +
h_{\nu}(-1+S_{\nu+1})}{2} \quad . \nonumber  
\eea
If we repeat these steps for the last term on the right hand side and
recall that by definition $S_{L-1}=0$, one ends up with
\bea{apdf}
& & h_{\nu}(S_{\nu}) \nonumber \\
&=& \sum_{n=0}^{L-2-\nu} \sigma^{(n+\nu+1)} \frac{1}{2^{n+1}}
\sum_{k=0}^{n+1} \left[ \left( \begin{array}{c} n \\ k \end{array} \right)
- \left( \begin{array}{c} n \\ k-1 \end{array} \right) \right]
h_{\nu}(n+1-2k + S_{n+1+\nu}) \nonumber \\
&+& \frac{1}{2^{L-1-\nu}} \sum_{k=0}^{L-1-\nu}
\left( \begin{array}{c} L-1-\nu  \\ k \end{array} \right) h_{\nu}(L-1-\nu-2k)
\quad .
\eea
Insert eqs.(\ref{apdf}) and (\ref{apdd}) into eq.(\ref{apdb}) and
equate the coefficients of $\sigma^{(\nu)}$ on both sides 
to obtain finally
\be{apdg}
\frac{2J}{L} S_{\nu} + H =
- g_{\nu}(S_{\nu}) - \sum_{\rho=1}^{\nu} \frac{1}{2^\rho}
\sum_{k=0}^{\rho} 
\left[ \left( \begin{array}{c} \rho-1 \\ k \end{array} \right)
- \left( \begin{array}{c} \rho-1 \\ k-1 \end{array} \right) \right]
h_{\nu-\rho}(\rho-2k + S_{\nu}) \quad .
\ee
This system of equations defines recursively the functions $g_{\nu}(S_{\nu})$
for increasing index $\nu$. Since the independent variable is discrete and
takes only a finite number of values, the numerical evaluation is quite
straightforward. It should be remarked that a simple asymptotic 
evaluation for large system size $L$ is possible too. 
{  Such an evaluation indicates that $g_{\nu}(S_{\nu})$ remains uniformly
bounded in the limit of infinite system size.}

\end{document}